\documentclass[runningheads]{llncs}
\usepackage[T1]{fontenc}
\usepackage{graphicx}
\usepackage{cite}
\usepackage{amsmath,amssymb,amsfonts}
\usepackage{cleveref}
\usepackage{algorithmic}
\usepackage{graphicx}
\usepackage{textcomp}
\usepackage{xcolor}
\usepackage{tikz}
\usepackage[style=british]{csquotes}
\usepackage{subcaption}

\usetikzlibrary{shapes.geometric, arrows}

\tikzstyle{startstop} = [rectangle, rounded corners, 
minimum width=3cm, 
minimum height=1cm,
text centered, 
draw=black, 
fill=red!30]

\tikzstyle{io} = [trapezium, 
trapezium stretches=true, 
trapezium left angle=70, 
trapezium right angle=110, 
minimum width=3cm, 
minimum height=1cm, text centered, 
draw=black, fill=blue!30]

\tikzstyle{process} = [rectangle, 
minimum width=3cm, 
minimum height=1cm, 
text centered, 
text width=3cm, 
draw=black, 
fill=orange!30]

\tikzstyle{decision} = [diamond, 
minimum width=3cm, 
minimum height=1cm, 
text centered, 
draw=black, 
fill=green!30]
\tikzstyle{arrow} = [thick,->,>=stealth]

\tikzstyle{box} = [rectangle, minimum width=3cm, minimum height=1cm,text centered, draw=black]

\def\BibTeX{{\rm B\kern-.05em{\sc i\kern-.025em b}\kern-.08em
    T\kern-.1667em\lower.7ex\hbox{E}\kern-.125emX}}

\begin{document}
\title{Investigating Matrix Repartitioning to Address the Over- and Undersubscription Challenge for a GPU-based CFD Solver}
%
%
\author{Gregor Olenik\inst{1}\orcidID{0000-0002-0128-3933} \and
Marcel Koch\inst{2}\orcidID{0009-0004-8333-9313} \and
Hartwig Anzt\inst{1}\orcidID{0000-0003-2177-952X}}

\authorrunning{G. Olenik et al.}
%
\institute{Chair of Computational Mathematics, TUM School of Computation, Technical University of Munich, Heilbronn, Germany\\
\email{\{gregor.olenik,hartwig.anzt\}@tum.de}\\
 \and
Scientific Computing Center, Karlsruhe Institute of Technology, Karlsruhe, Germany\\
\email{marcel.koch@kit.edu}}
\maketitle              
\begin{abstract}
Modern high-performance computing (HPC) increasingly relies on GPUs, but integrating GPU acceleration into complex scientific frameworks like OpenFOAM remains a challenge. Existing approaches either fully refactor the codebase or use plugin-based GPU solvers, each facing trade-offs between performance and development effort. In this work, we address the limitations of plugin-based GPU acceleration in OpenFOAM by proposing a repartitioning strategy that better balances CPU matrix assembly and GPU-based linear solves. We present a detailed computational model, describe a novel matrix repartitioning and update procedure, and evaluate its performance on large-scale CFD simulations. Our results show that the proposed method significantly mitigates oversubscription issues, improving solver performance and resource utilization in heterogeneous CPU-GPU environments.

\end{abstract}

\section{Introduction \label{sec:intro}}
Modern general-purpose GPUs have become an integral part of most HPC clusters. However, in many cases, adopting scientific research software to leverage GPU acceleration is an ongoing effort. One example is the widely used computational fluid dynamics (CFD) framework OpenFOAM \cite{jasak2007openfoam}. To this date, despite multiple efforts, no official OpenFOAM version has been released that provides general GPU support. Currently, two orthogonal approaches are pursued in the wider community: (1) a general refactoring of the OpenFOAM code base and (2) plugin-based approaches providing access to GPU capable linear solver.
Examples of the refactoring approach are 
OpenFOAM\_HMM \cite{tandon2024porting}, or zeptoFOAM. 
Furthermore, examples of the plugin-based approach are OGL \cite{Olenik_Towards_a_platformportable_2024} and petsc4FOAM \cite{bn2020petsc4foam}.
For a more comprehensive overview of projects that target OpenFOAM, the interested reader is referred to \cite{Olenik_Towards_a_platformportable_2024}. The re-implementation/refactoring approach promises larger performance benefits because here the complete simulation workflow including matrix assembly can be offloaded to the GPU and thus reduces the inherent communication overhead between host and device present. However, the refactoring approach has considerable development costs. Compared to the refactoring approach, the plugin approach has two main drawbacks, first, due to the design of OpenFOAM plugins, it can only offload the linear solver; consequently, matrix assembly remains on the CPU. This limits performance benefits to cases where matrix assembly contributes only a small share of the total computational costs. Second, if distributed computations are performed, finding an optimal partitioning of the computational domain such that matrix assembly on the CPU and linear solver on the GPU costs are well-balanced, is challenging. 
The authors of \cite{Mills2021} refer to this as the over- and under-subscription challenge. In the case of an OpenFOAM simulation with a GPU solver plugin, this can be illustrated as follows: consider a typical HPC node with two CPU sockets each with 64 cores per socket and four accelerators. Common decomposition strategies are to decompose the case into (i) $64 = N_{CPU}$ or (ii) $4 = N_{GPU}$ subdomains. Case (i) will ensure optimal performance for the matrix assembly portion of the computational cost, which generally benefits from more parallelism. However, with a na\"ive implementation, case (i) will also cause oversubscription of the GPUs with $N_{CPU}/N_{GPU}$ ranks per GPU. Especially, OpenMPI is very sensitive to oversubscribing GPUs, which can cause serious performance degradation \cite{bierbaum2022towards}. Additionally, an overhead is introduced from the additional communication required between ranks on the same device. Case (ii) avoids oversubscription of the GPUs by partitioning into $N_{GPU}$ subdomains, which additionally reduces the unnecessary communication between inter-device ranks by reducing the number of cells at processor boundaries on the same GPU. The drawback, however, is that matrix assembly, which is performed on the CPU, can only utilize  $N_{GPU}$ CPU cores, leading to an under-subscription of the CPU. While an optimal partitioning can be found based on a computational costs model that considers the computational costs of the matrix assembly, the linear solve, and the expected speed-up when employing more parallelism, it is a substantial change from the typical \enquote{use all available cores} workflow. 
In this work, we investigate a repartitioning approach to mitigate the drawbacks of case (i) when utilizing as many CPU cores as possible. The rest of this manuscript is structured as follows: in \cref{sec:CompApproach} a brief discussion of the underlying computational procedure and a basic computational cost model is presented, \cref{sec::RepartProcedure} presents a detailed discussion of how matrices on the CPU (host) and GPU (device) side are generated and updated, and finally \cref{sec:results}  presents performance data of the implemented repartitioning procedure for a common benchmark case.

\section{Computational Cost Model\label{sec:CompApproach}}

\begin{figure}[tbp]

\begin{center}
\begin{tikzpicture}[node distance=1.5 cm,thick,scale=1, every node/.style={scale=0.75}]]
\node (start) [startstop] {Begin timestep};
\node (mompredass) [process, below of=start] {Assemble momentum predictor};
\node (mompredsolv) [process, below of=mompredass] {Solve momentum predictor};

\node (dec1) [decision, below of=mompredsolv, yshift=-1.0cm, align=center] { PISO\\converged};

\node (corass1) [process, below of=dec1, yshift=-1.0cm] {Assemble pressure corrector};

\node (corsolve) [process, below of=corass1] {Solve pressure corrector};

\node (cormom) [process, right of=dec1, xshift=3.0cm] {Correct momentum solution};
\node (stop) [startstop, below of=cormom, yshift=-0.5cm] {End of timestep};

\draw [arrow] (start) -- (mompredass);
\draw [arrow] (mompredass) -- (mompredsolv);
\draw [arrow] (mompredsolv) -- (dec1);
\draw [arrow] (corsolve) -- ++(-2.0,0) |-  (dec1);
\draw [arrow] (dec1) -- node[anchor=east] {no} (corass1);
\draw [arrow] (dec1) -- node[anchor=south] {yes} (cormom);
\draw [arrow] (cormom) -- (stop);
\draw [arrow] (corass1) -- (corsolve);
\end{tikzpicture}
\end{center}
\caption{Flow chart of the principal steps within a timestep in the 
icoFOAM solver.\label{fig:flowchart}}
\end{figure}
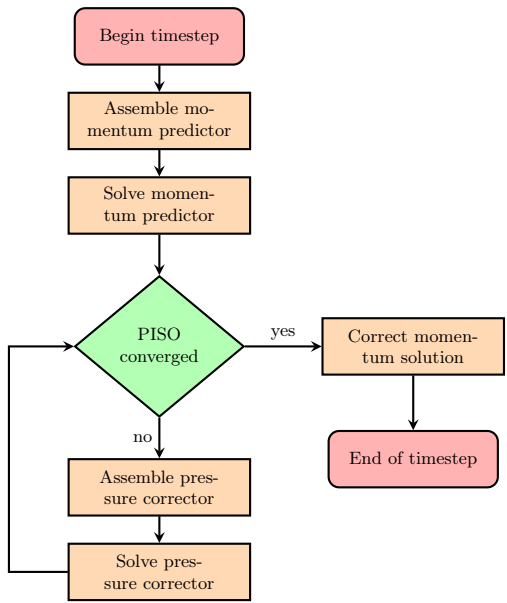

The OpenFOAM framework implements multiple finite volume method (FVM) based computational fluid dynamics (CFD) solvers, which typically are based on segregated projection methods like SIMPLE, PISO, or PIMPLE. As a consequence, linear systems of multiple partial differential equations, namely the momentum and the pressure equations, are assembled into separate linear equations and are usually solved in an iterative way. The computational procedure of OpenFOAM's icoFOAM solver is illustrated in \cref{fig:flowchart}. Within each time step, the set of linear equations for the momentum equation is assembled and solved. The result is used as a predictor of the velocity field and serves as an input for the PISO procedure, which assembles and solves the linear equations for a pressure corrector in an iterative way. After reaching the final iteration of the PISO loop, the velocity solution gets corrected by applying the effect of the pressure corrector to yield a conservative velocity field.
It should be noted that the icoFOAM solver serves as a model solver in this work, neglecting, for example, other  computational costs like non-orthogonal correctors, IO, or additional physical models. However, the proposed procedure can be beneficial to other solvers as well. 
Neglecting any additional costs, the time required to solve a single time step $T$ is the sum of the total matrix assembly and total linear solver costs, written as
\begin{align} \label{eq:Tn}
    T(n) = T_{AS}(n) + T_{LS}(n),
\end{align}
where $n$ is the number of used MPI ranks.
Each part of the RHS in \cref{eq:Tn} could be split further into individual equations depending on the OpenFOAM solver, e.g. the momentum and pressure for icoFOAM.
In the following, we assume that the individual equations have the same speed-up regarding the number of used MPI ranks, thus splitting each part into equations is not considered further.
The speed-up per part is defined as $S_{LS}(n) = T_{LS}(1) / T_{LS}(n)$, and $S_{AS}(n)$ accordingly.
In the heterogeneous setting, i.e. employing a different number of MPI ranks for matrix assembly and the linear solver, the speed-ups behave differently, that is $S_{LS} \neq S_{AS}$.
This could be the case, for example, if the assembly is done on CPUs, while the linear solver is offloaded to GPUs, which matches the OpenFOAM approach as discussed in \cref{sec:intro}.

The speed-ups attain their maximum at different inputs $N_{LS}^* < N_{AS}^*$ with $S_{LS}(n) \leq S_{LS}(N_{LS}^*)$, and $S_{AS}(n) \leq S_{AS}(N_{AS}^*)$.
In the case of ideal scaling, the optimal number of MPI ranks $N^*$ are the maximal available resources.
Thus, $N_{AS}^* = N_{CPU}$ is the number of available CPUs, and $N_{LS}^* = N_{GPU}$ is the number of available GPUs.
Considering the total runtime again
\begin{align}
    T(n) &= T_{AS}(n) + T_{LS}(n) \nonumber\\
         &= \frac{T_{AS}(1)}{S_{AS}(n)} + \frac{T_{LS}(1)}{S_{LS}(n)}
\end{align}
neither $n = N_{AS}^*$, nor $n = N_{LS}^*$ will achieve the minimum runtime.
Instead, optimizing the runtime requires some value for $n$ which achieves suboptimal speed-ups in both components.
Thus, some computing resources must be left unused.

However, by choosing the number of MPI ranks for $T_{AS}$ and $T_{LS}$ independently, optimal utilization for both components can be achieved.
Then the runtime can be expressed as
\begin{align}
    T(n_{AS}, n_{LS}) = T_{AS}(n_{AS}) + T_{LS}(n_{LS}) + T_R(n_{AS}, n_{LS})
\end{align}
where $T_R$ contains the cost to communicate between the two different MPI groups.
Now, it is clear that using the pair $(N_{AS}^*, N_{LS}*) = (N_{CPU}, N_{GPU})$ minimizes the cost for both the assembly and linear solver component.
Only the cost for the $T_R$ term has to be accounted for.
Additionally, this approach can be easily integrated into existing workflows, as already existing decompositions into $N_{CPU}$ can be reused.

\section{The Repartitioning Procedure \label{sec::RepartProcedure}}
In the following, we will use the term repartitioning for the process of mapping from a CPU partition, used for assembling the linear system, to a separate GPU partition, used to solve the linear system. 
Each partition consists of a number of parts, which are assigned an MPI rank $r$.
The number of parts in the CPU partition $n_{CPU}$ is determined by the domain decomposition procedure, which is applied as a preprocessing step by the utility \verb|decomposePar| that is part of the OpenFOAM framework.
In contrast, the number of parts in the GPU partition $n_{GPU}$ depends on a repartitioning ratio $\alpha \geq 1$ and is determined at run time
as $n_{GPU} = n_{CPU}/\alpha$, i.e. it defines the number of ranks per GPU.
Thus, for the number of parts the following relationship holds: $n_{GPU}\leq n_{CPU}$.
To facilitate the mapping between the two partitions, connections between ranks in the CPU partition and ranks in the GPU partition need to be defined.
The connection defines which degrees of freedom (DOF) in the linear system are owned by which ranks.
In this work, a blockwise distribution is considered, where the GPU rank $k$ owns the same DOFs as the $\alpha$ CPU ranks $\{\alpha k, \alpha k + 1, \dots, \alpha k + \alpha - 1\}$.

For both the CPU and GPU partitions, the matrix coefficients are stored according to their communication properties.
That is, all matrix coefficients that are applied solely to local components are stored in a local matrix and coefficients requiring components from a different rank are stored separately in a non-local matrix.
An additional constraint is given by the fact that in OpenFOAM, the host matrix is stored in LDU format. Thus, additionally, a mapping between LDU matrix entries and the entries of device matrix format is required.

The repartitioning procedure investigated for this work can be described as:
\begin{enumerate}
\item Extract the sparsity pattern from the host matrix, including all coupling terms with non-local entries.
\item Send local and non-local sparsity patterns from CPU ranks to their connected GPU ranks.
\item On the GPU ranks, all received local sparsity patterns are fused into a single local sparsity pattern, i.e. the GPU rank $k$ fuses the sparsity pattern received from the CPU ranks $\{\alpha k, \alpha k + 1, \dots, \alpha k + \alpha - 1\}$. The received non-local matrix elements are included in the new local sparsity pattern, if the former communication partner resides on the same part after repartitioning. Otherwise, they are fused in the non-local matrix.
\end{enumerate}

\begin{figure}[h!]
\centerline{\includegraphics[width=\linewidth]{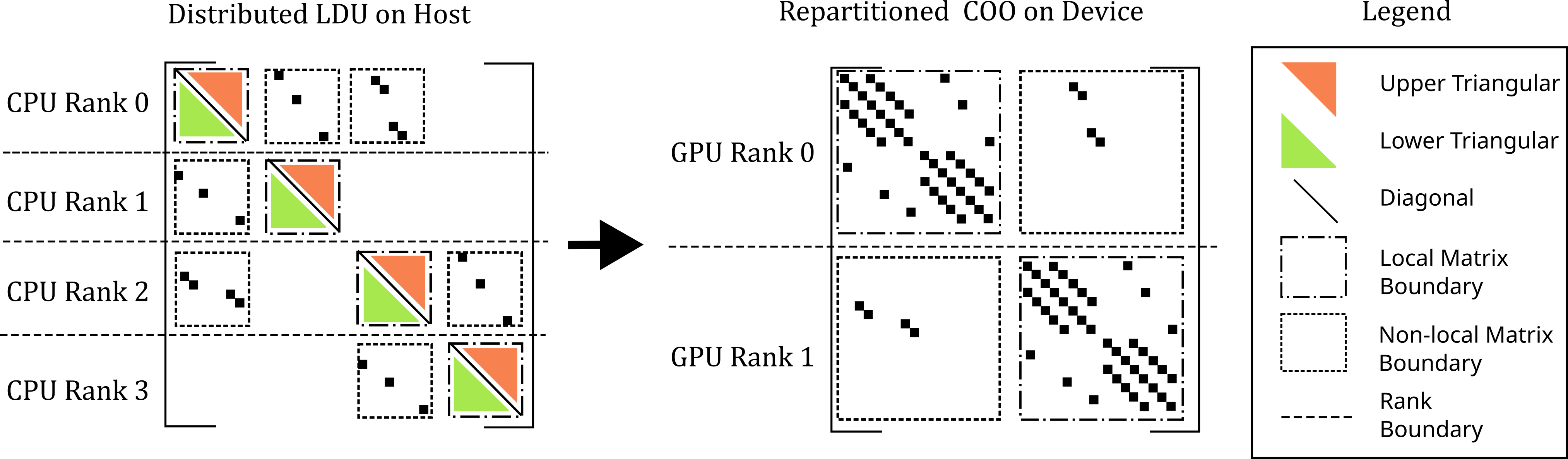}}
\caption{Structure of the distributed matrix in LDU format on the host (top) and after repartitioning on the accelerator (bottom), with $\alpha = 2$.}
\label{fig:matrix_structure}
\end{figure}

To illustrate, \Cref{fig:matrix_structure} shows the initial structure of the distributed LDU matrix on the host (top) and the resulting matrix in COO format on the accelerator after repartitioning (bottom). 
By reducing the number of individual parts, the resulting repartitioned matrix also has fewer coefficients in the non-local interface matrix. 

\begin{figure*}[t!]
    \centering
    \begin{subfigure}[t]{0.48\textwidth}
        \centering
        \includegraphics[width=0.9\textwidth]{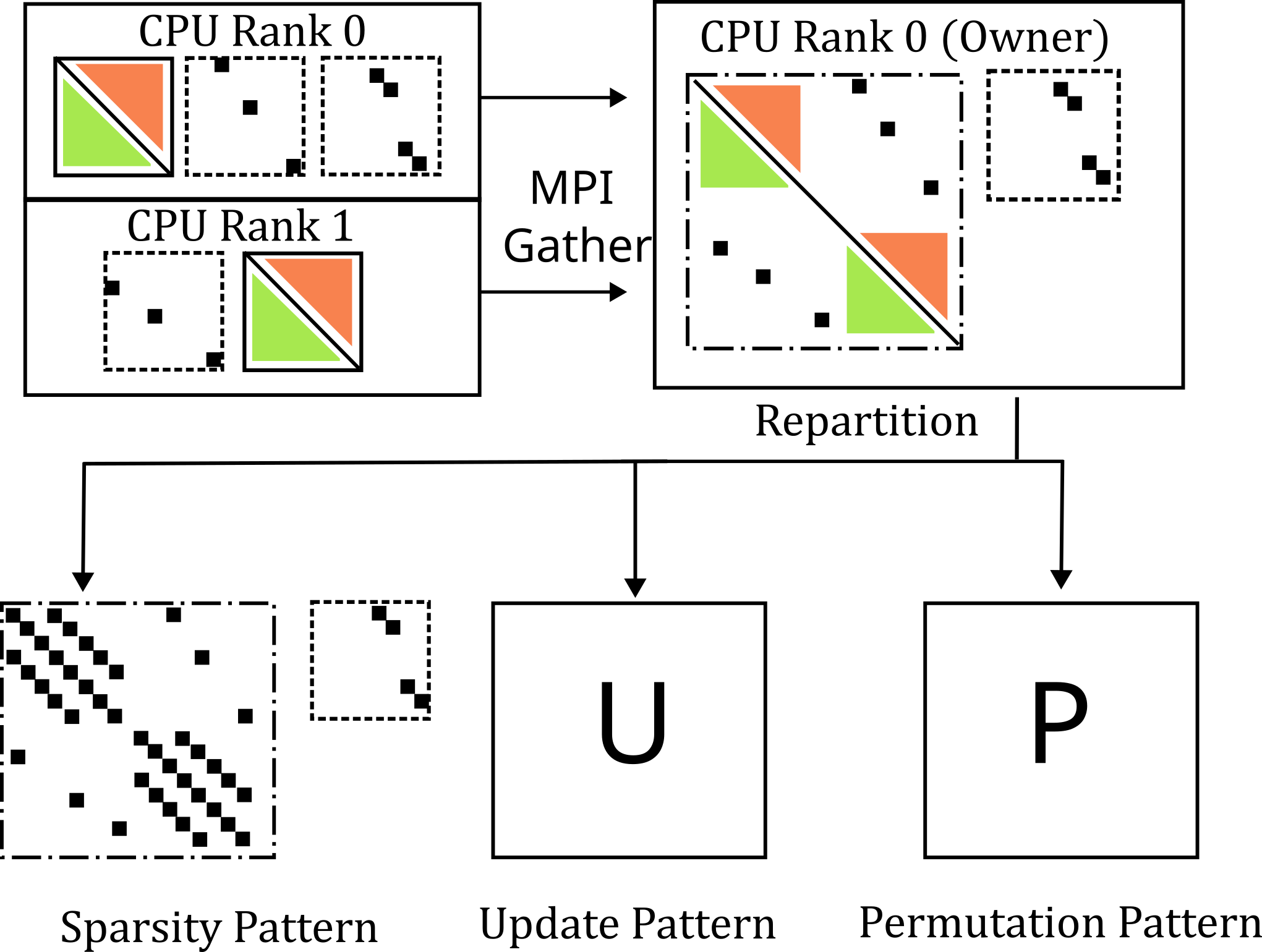}
        \caption{Resulting data structures to for updating of the matrix coefficients.\label{fig:repart_procedure}
    }
    \end{subfigure}%
    ~ 
    \begin{subfigure}[t]{0.48\textwidth}
        \centering
        \includegraphics[width=0.9\textwidth]{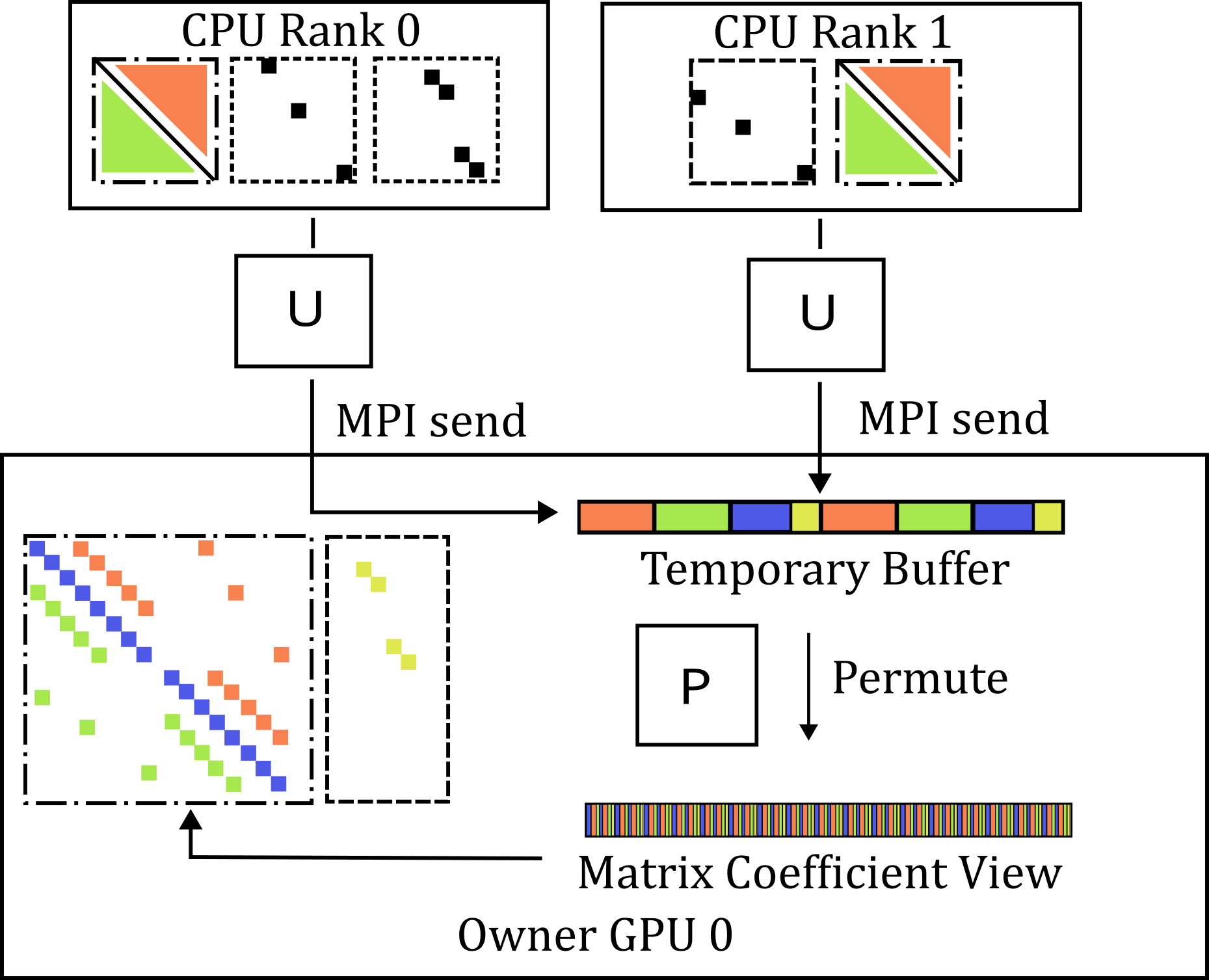}
        \caption{Update procedure of the distributed matrix coefficients.\label{fig:repart_update}}
    \end{subfigure}
    \caption{Illustration of the repartitioning procedure.}
\end{figure*}

The approach presented first applies the repartitioning procedure to the sparsity pattern of the matrix. Hence, it yields a repartitioned matrix without computing the repartitioned matrix coefficients.
Thus, in a separate step, the matrix coefficients of the repartitioned matrix have to be set on the basis of the matrix values of the host matrix.
From step (3) a mapping between the ordering of the host matrix in LDU format and the required ordering of the matrix on the destination rank, e.g. a row-major ordered COO matrix, is created. To update the matrix coefficients, each rank sends its matrix coefficients to the corresponding owner GPU rank into a continuous buffer array. The owning rank then orders the data according to the computed mapping into a view of repartitioned device matrix.
The principal process of repartitioning the matrix coefficients is illustrated in \cref{fig:repart_procedure}. 
The repartitioner combines several CPU-based LDU blocks by gathering onto an owning GPU MPI rank. 
As mentioned before, the assignment from CPU-to-GPU rank follows block-wise. Other methods for distributing matrix elements among parts that might yield more optimal partitions, e.g. Metis, are possible but lie outside the scope of this work.
Additionally, the repartitioner is responsible for localizing non-local blocks that have purely local communication after repartitioning. 
These can be identified as blocks in which global column index $j$ is within the set of global row indices $I_{GPU}(r)$ of the fused sparsity pattern on the GPU with rank $r$ i.e. $j \in I_{GPU}(r) = \bigcup_{l=0}^{\alpha - 1} I_{CPU}(\alpha r + l)$.

The process of repartitioning yields three data structures:
\begin{enumerate}
\item The sparsity pattern, i.e. row and column indices, of the repartitioned matrix.
\item An update pattern $U$ for the repartitioned coefficients. Here, the update pattern stores on each rank (a) the target ranks of its send operation, (b) send and receive pointers to the host data and the target data buffer respectively, and (c) the corresponding sizes for the MPI communication. This allows each rank to send its corresponding LDU matrix data to a designated buffer on the device on the receiving owner rank.
\item A permutation matrix which maps from the original LDU-based ordering to a row-major ordering of the resulting repartitioned device matrix.
\end{enumerate}

To realize the different CPU and GPU partitions, a new MPI communicator is created by splitting the CPU MPI communicator $C$ into active $C_{a}$ and inactive $C_{i}$ ranks, where $C_{a}$ contains the active ranks in the GPU partition.
The default communicator $C$ is required for matrix coefficient updates.
The communicator $C_{a}$ is passed to the linear solver handling the distributed linear system, while the inactive ranks $C_i$ just skip this step.
The communicator $C_{a}$ is then used exclusively on the GPUs.
This process avoids creating any empty matrices on devices for non-owning ranks, since this can cause significant performance degradation.

To ensure optimal performance, creating and updating the system matrix are two distinct processes.
This allows one to create the repartitioned, distributed matrix once and reuse it by updating its value through the course of a simulation.
The update procedure for the matrix coefficients is illustrated in \cref{fig:repart_update}.
If GPU aware MPI is available, each host rank sends its local data directly to a memory location of a temporary buffer on the device. Otherwise, the data is gathered on the CPU ranks $\alpha k$ first and then copied to the GPU in a separate step. This results in an array that is consecutively ordered regarding the original LDU order and the originating ranks.
Thus, in a second step, the coefficients are reordered by the permutation matrix $P$ to satisfy the row-major ordering expected by the linear solver library.

\section{Performance Evaluation \label{sec:results}}
\begin{figure}[htbp]
\centerline{\includegraphics[height=0.33\linewidth]
{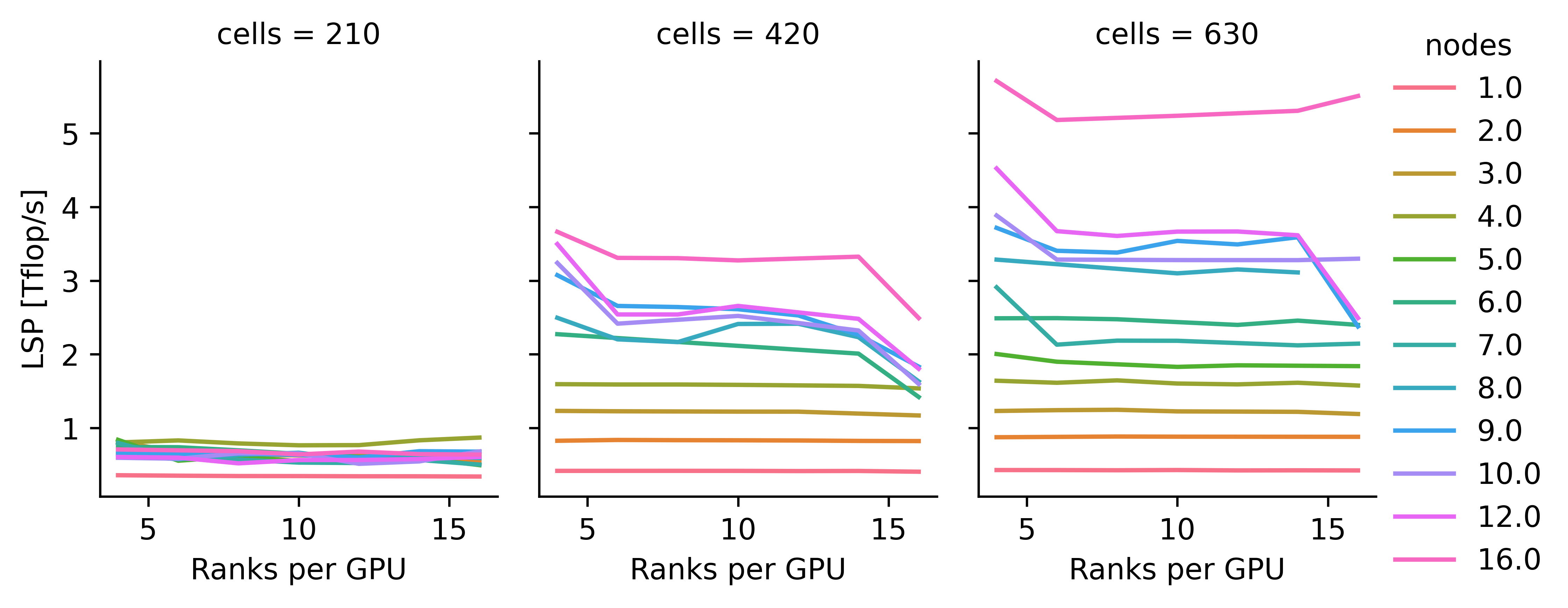}
}
\caption{Impact of the repartitioning ratio RPG on the linear solver performance in Tflop/s for a different number of compute nodes and problem sizes.}
\label{fig:rpg_performance}
\end{figure}

For the performance evaluation, the GPU based linear solvers of the Ginkgo library \cite{GinkgoJoss2020}  are employed. The Ginkgo library is made available to OpenFOAM via the OpenFOAM-Ginkgo Layer (OGL) \cite{Olenik_Towards_a_platformportable_2024} plugin. The repartition procedure discussed in \Cref{sec::RepartProcedure} is fully implemented within OGL.
The lidDrivenCavity3D case \cite{bn2020petsc4foam} is taken as a benchmark case. 
The computational grid is based on a uniform cubic grid with
$2\times 3 \times 5 \times 7 \times n_{p}$ cells along each axis, where $n_{p}$ is the factor controlling the overall size of the problem.
This ensures that the resulting computational domains are decomposable into equally sized subdomains by a wide range of factors. The value of $n$ is $1$, $2$, or $3$ for the small, medium, and  large cases with approximately 9M, 74M, and 250M cells, respectively. The time steps of the corresponding runs are adapted to ensure a constant CFL number. The cases are decomposed into subdomains using the OpenFOAM multilevel decomposition strategy, with a simple decomposition on the outermost node level and Scotch for the GPU and CPU levels, with $n_{tot, domains} = n_{nodes} \times n_{GPUs} \times \alpha$. The pressure equation is solved by Ginkgos CG solver for the GPU runs and with OpenFOAMs PCG for the CPU reference case. For solving the momentum equation, which accounts only for a small portion of the total compute time, OpenFOAMs native BiCGStab solver is used. 
The cases are run for exactly 20 time steps to ensure a balance between reasonable computational costs and sufficient computational work to extract meaningful statistics. For statistical analysis, an average of all computational time steps is calculated, excluding the first time step. The setup, execution and post-processing of the parameter study conducted is handled by the OpenFOAM Benchmark Runner (OBR) \cite{Gaertner_2025}.
The numerical experiments were conducted on the HoreKa HPC cluster, which is equipped with two Intel Xeon Platinum 8368 CPUs and four NVIDIA A100-40 GPUs per compute note. Furthermore, the software stack included the following software versions: OpenFOAM 2412, Ginkgo 1.9, OpenMPI 5.0.1, with UCX v1.18.0 and CUDA 12.2. 
\Cref{fig:rpg_performance} shows the impact of different repartitioning ratios $\alpha$ on the linear solver performance (LSP), reported as TFLOP/s. It can be seen that the performance of the linear solver is mostly independent of the repartitioning ratio $\alpha$ and is mainly influenced by the problem size and the number of compute nodes used.
However, in some cases, performance degradation was observed. This typically occurred when more than one, four, or six compute nodes were used for the small, medium, and large case, respectively. This indicates
that a minimum number of DOFs per GPU device is required for a constant performance across all repartitioning ratios. 

\begin{figure}[h!]
\centerline{\includegraphics[height=0.33\linewidth]
{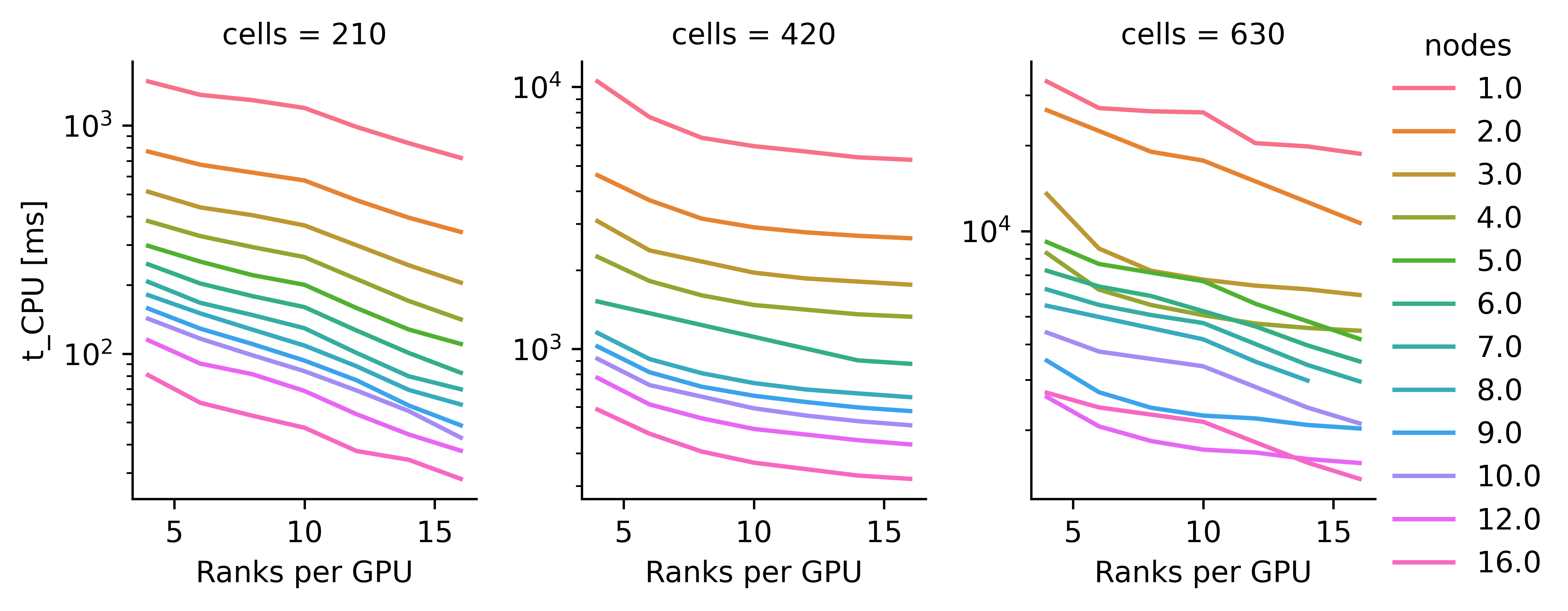}
}
\caption{Impact of the repartitioning ratio RPG on the time spent on host-side computations for different problem sizes and number of compute nodes.}
\label{fig:host-comp}
\end{figure}

\begin{figure}[h!]
\centerline{\includegraphics[height=0.33\linewidth]{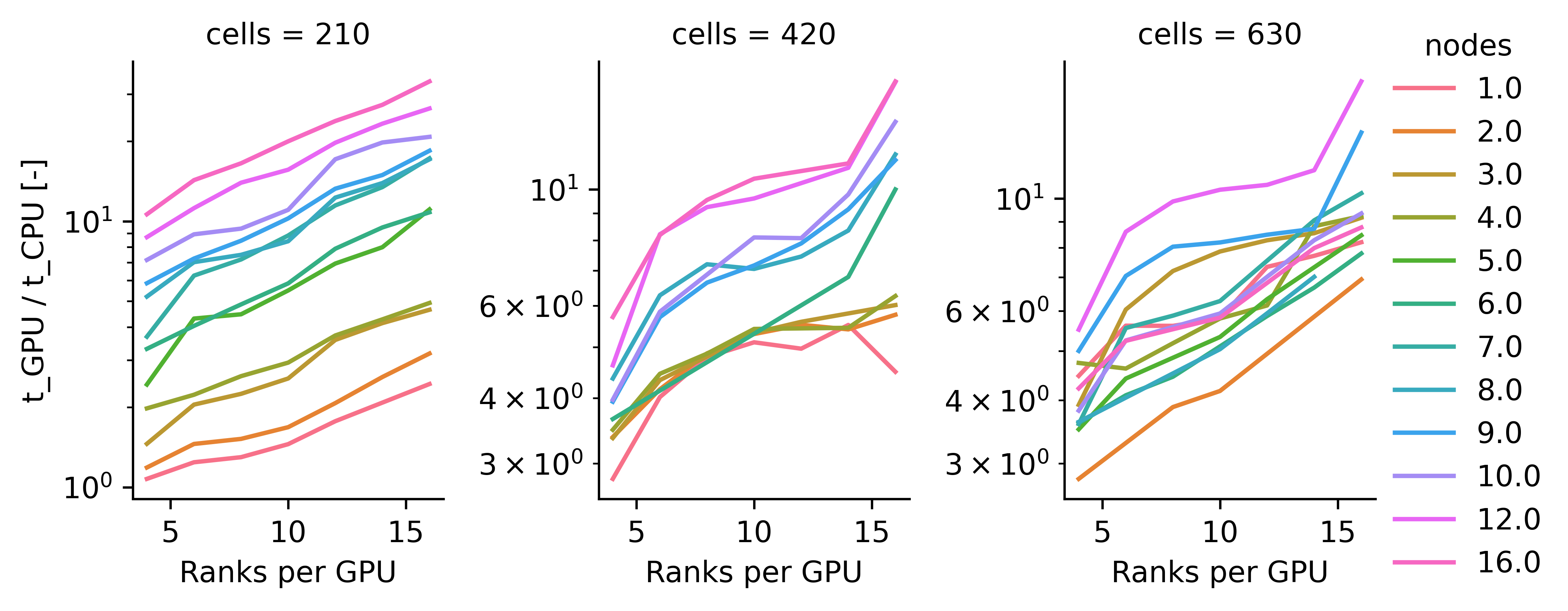}}
\caption{Impact of the repartitioning ratio $\alpha$ on the ratio of time spent on the GPU to CPU for different problem sizes and number of compute nodes.}
\label{fig:device-host-ratio}
\end{figure}

While the performance of linear solver is approximately constant for different repartitioning ratios, an increasing $\alpha$ allows employing more MPI-Ranks for computations residing on the host, e.g. the  matrix assembly. Thus, employing more parallelism on the host side should speed up the host-side computations. The time spent on the host side computations is given in \cref{fig:host-comp}, which shows a clear trend of decreasing the time required on the CPU with increasing $\alpha$.
With a decreasing amount of time spent on the host side computations, the ratio $\phi=t_{GPU}/t_{CPU}$ between the time spent on the GPU device and the CPU increases 
as shown in \cref{fig:device-host-ratio}. Thus, for a sufficiently large number of compute nodes and $\alpha$, the time spent on the host side becomes almost negligible as $phi$ approaches values between 15 and 30.
After ensuring a reasonable behavior of the repartitioning strategy over a wide set of parameters and problem sizes, a limited number of cases were selected to study the impact of the repartitioning strategy compared to typical alternative partitioning strategies.  

\begin{figure}[h!]
\centerline{\includegraphics[height=0.33\linewidth]{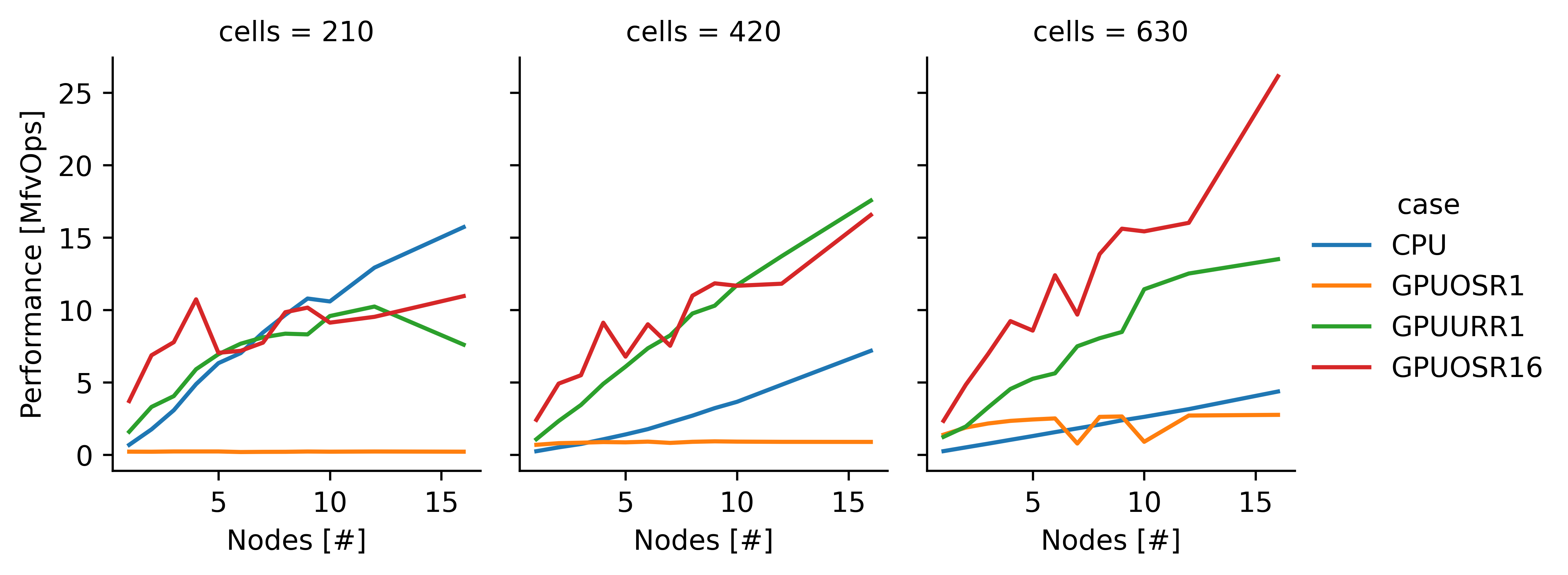}}
\caption{Strong scaling performance for the different partitioning strategies, undersubscribing (GPUURR1), oversubscribing (GPUOSR1), repartitioning (GPUOSRR16) compared to the unaccelerated reference case (CPU)  \label{fig:Mfvops}}
\end{figure}

\begin{figure}[h!]
\centerline{\includegraphics[height=0.33\linewidth]{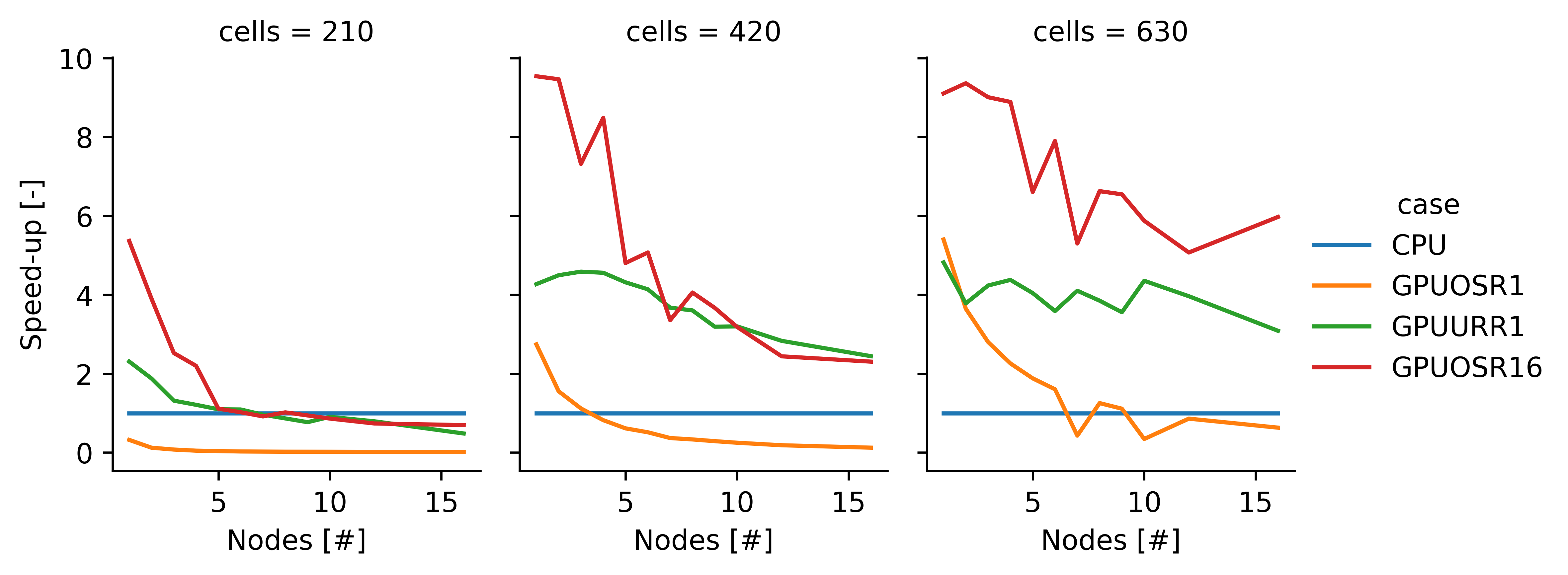}}
\caption{Speedup of the tested partitioning strategies: undersubscribing (GPUURR1), oversubscribing (GPUOSR1), repartitioning (GPUOSRR16) wrt. to the reference case (CPU) for different problem sizes and number of compute nodes.}
\label{fig:Speedup}
\end{figure}

\begin{figure}[h!]
\centerline{\includegraphics[height=0.33\linewidth]{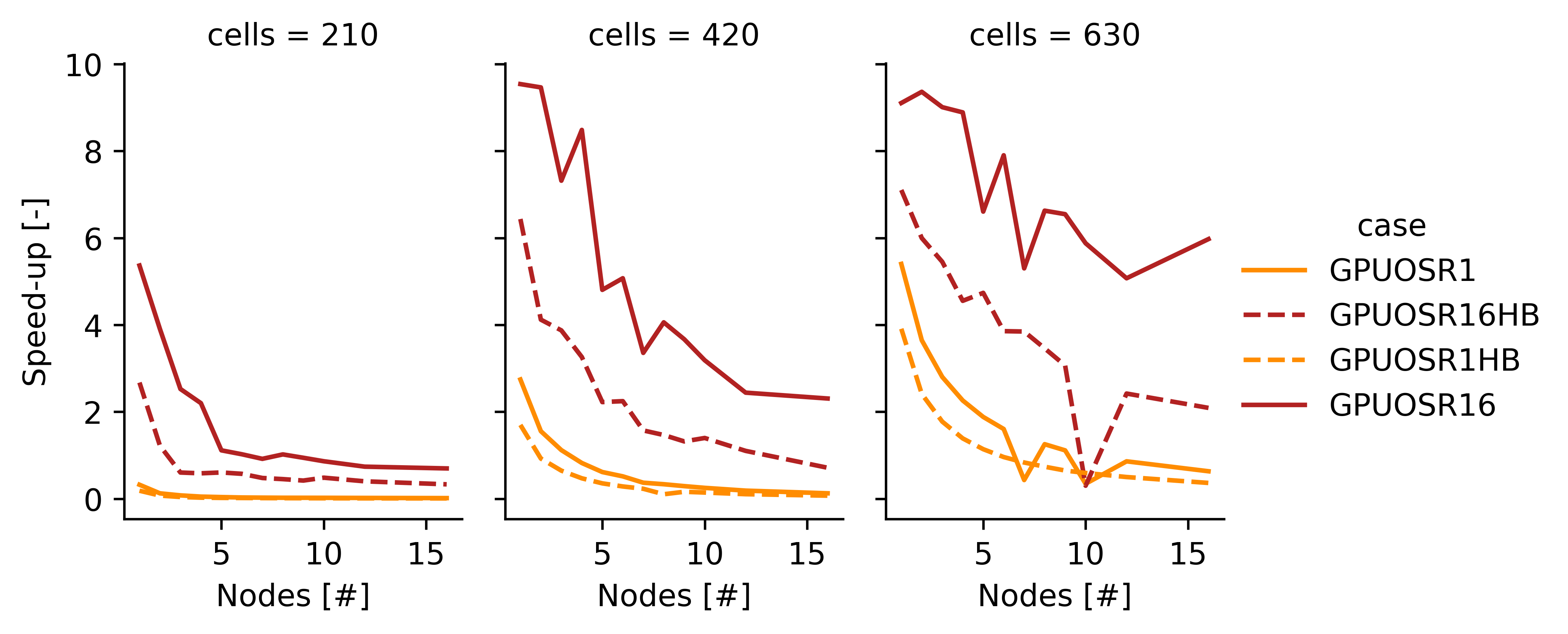}}
\caption{Impact of GPU aware MPI functionalities (solid lines) compared to communication via a host-side buffer (dashed lines and HB suffix in the case name) for different partitioning strategies, problem sizes, and number of compute nodes. \label{fig:GPUAwareMPI}}
\end{figure}

Thus, for further investigation of the impact of offloading and repartitioning on the performance PISO algorithm shown in \cref{fig:Mfvops} the following cases were selected for further investigation : (1) the reference case (blue line) without GPU acceleration using all available CPU cores, (2) a repartitioned case (red line) where 16 CPU MPI ranks were mapped to one GPU rank. (3) an unpartitioned case (green line, GPUURR1) in which exactly $n_{GPU}$ MPI ranks were used for linear solvers and matrix assembly and (4) an unrepartitioned case (orange line, GPUOSR1) that uses $n_{CPU}$ ranks and thus overutilizes the GPUs.
\Cref{fig:Mfvops} shows the strong scaling behavior performance $P=n_{cells}/t_{TS}$ in $10^6$fvOps on up to 16 nodes with 64 GPUs. 
The following findings can be observed here; the compute performance increases with increasing problem size on the GPUs, from $420^3$ compute cells both cases GPUURR1 and GPUOSR16 perform consistently better compared to the unaccelerated cases. The GPUOSR1 case, on the other hand, shows a significant reduction in performance of up to a factor 140. It is therefore not advisable to use unrepartitioned and overloaded cases in combination with OpenMPI.
\Cref{fig:Speedup} compares the speed-up of the different partitioning strategies of the accelerated version regarding the non-accelerated CPU code for the small, medium, and large computational grid over different number of compute nodes. In general, it can be observed that the speed-up of the accelerated cases increases for increasing problem sizes and decreases for increasing number of nodes. Which, suggests that the accelerated cases profit from a larger number of DOFs per GPU in contrast to the CPU case. This is in line with the findings of \cite{galeazzo2024understanding} which suggest that OpenFOAM cases can profit from optimal cache utilization when partitioning approx. 10000-30000 DOFs per CPU core. In contrast, optimal utilization of the GPU is observed for more than 1M DOFs per GPU. This leads to better parallel efficiency of the CPU cases compared to the GPU cases when increasing the number of compute nodes and explains the decreasing speed-up with decreasing number of DOFs per compute unit. Of the tested cases, the repartitioned case with $\alpha=16$ (GPUOSRR16) performed best over a wide range of nodes and cases with a maximum acceleration of up to a factor of 10, however, for a larger number of compute nodes, the results are converging towards the unrepartitioned case GPUUSR1. Of the accelerated cases tested, the oversubscription case without repartitioning (GPUOSR1) showed the worst performance with a speed-up of 0.007 in the worst case, which seems to be a consequence of the aforementioned oversubscription issue of OpenMPI \cite{bierbaum2022towards}. 
In summary, the use of the newly developed repartitioning method suceeds in matching computing performance using a smaller number of computing nodes.
Furthermore, for the presented repartitioning method, efficient communication among CPU-GPU and GPU-GPU is crucial. The former is required for an efficient update of the matrix coefficients, and the latter is required for distributed SpMVs in the distributed linear solver. Thus, the impact of enabling GPU-aware MPI and GPU-direct is shown in \cref{fig:GPUAwareMPI} for the repartitioning case GPUOSRR16 and the oversubscribing case GPUOSR1. Here, solid lines indicate cases that take advantage of GPU-aware MPI by directly sending data to buffers on the GPU device.
For the cases without GPU aware MPI (dashed lines and HB suffix), the data were first copied from the device back to the host before communicating with other ranks. A substantial impact on the performance can be observed that ranged from 25\% to 50\%
\section{Conclusion and Outlook}
In this work, a repartitioning procedure was investigated to address the challenge of over- and undersubscription, which is a consequence of the heterogeneous computing approach of GPU offloading plugins for OpenFOAM. It was demonstrated that by repartitioning significant performance gains are achievable compared to the unaccelerated case. Additionally, it performed better against the accelerated cases with either undersubscribing or oversubscribing. The investigated lidDriven3D case spent the majority of the compute time for the linear solver. Thus, more complex test cases should be investigated, since for industrially relevant cases often an equal ratio between matrix assembly and linear solver costs is reported. Additionally,
 the impact of repartitioning on other solver like the distributed multigrid solvers and the convergence behavior of the preconditioner should be investigated. 

\section*{Acknowledgment}


This work was supported by the German Federal Ministry of Education and Research (grant number 16ME0676K). This work was performed on the HoreKa supercomputer funded by the
Ministry of Science, Research and the Arts Baden-Württemberg and by the Federal Ministry of Education and Research.

\bibliographystyle{splncs04}
\bibliography{references2}

\end{document}